\documentstyle[preprint,aps,prl,psfig]{revtex}
\tighten
\draft
\begin{document}

\draft


\title{The structural relaxation of molten sodium disilicate}
\author{J\"urgen Horbach$^1$ and Walter Kob$^2$\footnote{Work dedicated to the 60th
birthday of J.--P. Hansen}}
\address{$^1$ Institut f\"ur Physik, Johannes Gutenberg--Universit\"at,
Staudinger Weg 7, D--55099 Mainz, Germany\\
Laboratoire des Verres, Universit\'e Montpellier
II, 34095 Montpellier, France}
\maketitle

\begin{abstract}
We use molecular dynamics computer simulations to study the relaxation
dynamics of Na$_2$O--2(SiO$_2$) in its molten, highly viscous state. We
find that at low temperatures the incoherent intermediate scattering
function for Na relaxes about 100 times faster than the one of the
Si and O atoms. In contrast to this all coherent functions relax on
the same time scale if the wave--vector is around 1\AA$^{-1}$. This
anomalous relaxation dynamics is traced back to the channel--like
structure for the Na atoms that have been found for this system. We
find that the relaxation dynamics for Si and O as well as the time
dependence of the coherent functions for Na can be rationalized well
by means of mode--coupling theory. In particular we show that the
diffusion constants as well as the $\alpha$--relaxation times follow the
power--law predicted by the theory and that in the $\beta$--relaxation
regime the correlators obey the factorization property with a master
curve that is described well by a von Schweidler--law. The value of the
von Schweidler exponent $b$ is compatible with the one found for the
mentioned power--law of the relaxation times/diffusion constants. Finally
we study the wave--vector dependence of $f_s(q)$ and $f(q)$, the coherent
and incoherent non--ergodicity parameters. For the Si and O atoms these
functions look qualitatively similar to the ones found in simple liquids
or pure silica, in that the coherent function oscillates (in phase with
the static structure factor) around the incoherent one and in that the
latter is approximated well by a Gaussian function. In contrast to this,
$f(q)$ for Na--Na is always smaller than $f_s(q)$ for Na and the latter
can be approximated by a Gaussian only for relatively large $q$.

\end{abstract}


\section{Introduction}
\label{sec1}
In the last two decades our understanding of the structural
and dynamical properties of glass forming liquids has increased
impressively~\cite{heraclion,alicante,vigo,hernonissos}. This progress
is due to significant advances in various experimental techniques (light
and neutron--scattering, dielectric measurements, etc.), the development
of new theoretical approaches and concepts (mode--coupling theory (MCT),
landscapes, etc.)~\cite{mct,sastry98,sastry01,debenedetti01} and last
not least to the extraordinary advances that computer simulations have
made~\cite{angell81,barrat91,kob_stauffer95,poole95,kob99}. The results
of all these efforts is a widely accepted picture on the relaxation
dynamics of glass--forming liquids: At high temperatures this dynamics
is the one of a normal liquid and hence the typical time correlation
functions decay exponentially in time and the $T$--dependence of the
typical relaxation times $\tau(T)$ shows an Arrhenius--law. (Note
that although there is a well developed theoretical machinery to
describe the {\it static} properties of such liquids~\cite{hansen86},
an {\it accurate} understanding of the {\it dynamics} is still lacking.) With
decreasing temperature the dynamics changes in that time correlation
functions are no longer exponential and the $T$--dependence of the
relaxation times is often super--Arrhenius. In the past it has
been shown that the slow dynamics observed in this temperature
range can be rationalized very well by means of the so--called
mode--coupling theory of the glass transition, and that often this
theory is not only able to describe this dynamics qualitatively but even
quantitatively~\cite{mct,barrat89,megen94,nauroth97,fabbian99,theis00a,theis00b,winkler00}.
If the temperature is decreased even further the relaxation times
increase very quickly and typically show an Arrhenius dependence on $T$
with an activation energy that is larger than the one found at higher
temperatures. The details of the dynamics in this temperature range are
not understood very well and also there exist no satisfactory theoretical
description for it.

The presence of the three mentioned regimes in $\tau(T)$ (Arrhenius,
super--Arrhenius, Arrhenius) is the typical behavior found in the
so--called ``fragile'' glass--forming liquids~\cite{angell85}, which
include polymers, most molecular glass--formers, and simple liquids. In
contrast to this, the so--called ``strong'' glass--forming liquids (typical
examples include silica and many other oxide glasses) show in the whole
accessible temperature range an Arrhenius--law, i.e. no super--Arrhenius
temperature dependence is found. Since it is one of the main achievements
of MCT to rationalize this super--Arrhenius $T$--dependence, it was believed for
quite some time that the theory is not very useful for these type of
glass--formers. In recent computer simulations it was shown, however, that
most models for strong glass--forming liquids exhibit at sufficiently high
temperatures deviations from the Arrhenius law found at intermediate
and low temperatures~\cite{vollmayr96,hemmati98,horbach99,saika01,saika01b,saika02}
and that in the temperature range in which these deviations
are seen the relaxation dynamics can be described very well by
MCT~\cite{horbach99,sciortino01,horbach01b}, in agreement with the results
of some experiments~\cite{wuttke,franosch97}. Therefore one can conclude
that this theory is not only able to describe fragile glass--formers, but
also all the intermediate and strong ones.

The results discussed so far concern systems in which the dynamics of
all atomic species occurs on roughly the same time scale. There are,
however, large classes of materials in which this is not the case. E.g.~in
superionic glass--formers, such as Na$_2$O--$x$SiO$_2$ or mixed alkali
glasses such as 0.5Na$_2$O--0.5K$_2$O--3SiO$_2$, the alkali atoms move
on a time scale which at low temperatures is many orders of magnitude
faster than the one of the atoms constituting the matrix (here Si and
O)~\cite{jonscher77,ingram87,dyre91,kahnt96,funke99,dyre00}. To what
extent MCT is able to rationalize the very heterogeneous dynamics of these
type of glass--forming systems is presently unknown, although recently
evidence has been found that certain aspects of the relaxation dynamics
can indeed be understood by means of the theory~\cite{horbach02}. The
goal of the present paper is therefore to do a detailed investigation
of the relaxation dynamics of a prototype of such an ion--conducting
glass--former, Na$_2$O--2SiO$_2$, and to see to what extent the predictions
of MCT regarding the dynamics of glass--forming liquids hold. Note
that although so far there has been no test of MCT for these systems,
computer simulations have already been used for a long time to study such
ion--conducting materials. Seminal work on this goes back more than twenty
years to Soules {\it et al.}~\cite{soules79} who studied the structure
of sodium--disilicate and to Angell {\it et al.} who looked at certain
aspects of the diffusion dynamics~\cite{angell81b,angell82}. These
early investigations were more recently followed up by
similar studies on related systems and also many experimental
investigations~\cite{ingram89,huang90,vessal92,bala94,elliott94,smith95,sen97,greaves98,habasaki98,oviedo98,richert99,sidebottom99,schroder00,maass96,porto00,banhatti01,heuer02,zotov01,jund01,yuan01,sunyer02}.

A further interesting feature of these systems is that the distribution
of the alkali ions in the SiO$_2$ matrix is not completely uniform
but instead forms an interconnected network of small pockets and
filaments~\cite{ingram87,ingram89,smith95,oviedo98,jund01}. Although for
a long time there was no {\it direct} experimental evidence for the presence
of these structures, recent neutron--scattering experiments of Meyer {\it
et al.} on Na$_2$O--2SiO$_2$ ~showed a structural feature at a wave--vector
$q\approx 0.9$~\AA$^{-1}$~\cite{meyer02}, in excellent agreement with the
results from computer simulations that could attribute this peak to the
presence of the above--mentioned network~\cite{horbach02,horbach99a,horbach01}. How
the presence of these channels affects the relaxation dynamics is,
however, so far not known and therefore the present paper is also a
contribution to gain insight into this matter.

The remaining of the paper is organized as follows: In the next section
we will describe the model used for the simulations as well as give
their details. The following section is then devoted to the presentation
of the results and in the final section we summarize and discuss them.

\section{Model and Details of the Simulation}
\label{sec2}
The force field used in the present work is a modification of the one
proposed by Kramer~{\it et al.} to describe zeolites~\cite{kramer91},
i.e.~crystalline materials that contain also Si, O and Na. This potential
had the functional form

\begin{equation}
\phi_{\alpha\beta}(r)=
\frac{q_{\alpha} q_{\beta} e^2}{r} +
A_{\alpha \beta} \exp\left(-B_{\alpha \beta}r\right) -
\frac{C_{\alpha \beta}}{r^6} \quad \alpha, \beta \in
[{\rm Si}, {\rm Na}, {\rm O}],
\label{eq1}
\end{equation}

\noindent
where $r$ is the distance between two atoms of type $\alpha$ and
$\beta$. The parameters $A_{\alpha \beta}$, $B_{\alpha \beta}$, and
$C_{\alpha \beta}$ can be found in Ref.~\cite{kramer91}. Note that
$q_\alpha$ is an {\it effective} charge with values $q_{\rm Si}=2.4$
and $q_{\rm O}=-1.2$. In Ref.~\cite{kramer91} the effective charge of the
sodium atoms was chosen to be $q_{\rm Na}=1.0$, which has the disturbing
effect that Na$_2$O is not neutral. Hence we changed the potential of
Ref.~\cite{kramer91} somewhat, by assigning to the Na atom an effective
charge $q_{\rm Na}=0.6$~\cite{horbach01}. This change of the charge
affects of course the local structure and therefore we have added to
$\phi_{\alpha\beta}(r)$ a term which compensates this change at
{\it short} distances. Hence the potential we used in our simulation is
given by

\begin{equation}
\Phi_{\alpha\beta}(r)=\phi_{\alpha\beta}(r)+
\frac{\tilde{q}_i\tilde{q}_je^2}{r} \left[ 1- ( 1-\delta_{\alpha{\rm Na}}) 
( 1-\delta_{\beta{\rm Na}}) \right]\Theta(r_{{\rm c}}-r)
\label{eq2}
\end{equation}

\noindent
with $\tilde{q}_{\rm Si}=2.4$, $\tilde{q}_{\rm O}=-1.2$, and
$\tilde{q}_{\rm Na} = 0.6 \ln \left[ C (r_{\rm c}-r)^2+1 \right]$.
(Here $\Theta$ is the usual Heaviside function.) The parameters
$C=0.0926$~\AA$^{-2}$ and $r_{{\rm c}}=4.9$~\AA~were chosen such that
at ambient pressure the resulting structure is in good agreement with
the experimental results from neutron scattering~\cite{horbach01}. Thus
the potential $\Phi_{\alpha\beta}(r)$ used in the present work is at
short distances very similar to the one proposed by Kramer~{\it et al.}
and is at large distances (where the effective charges are screened)
modified such that all systems of the form Na$_2$O--$x$SiO$_2$ are neutral.

In previous studies it has been shown that this potential is able
to reproduce many structural properties of Na$_2$O--$x$SiO$_2$,
with $x=2,3,4$, and also certain aspects of the relaxation
dynamics of this model are in good qualitative agreement with
experiments~\cite{horbach02,jund01,horbach99a,horbach01,ispas01}. Although
it cannot be expected that such a simple model is able to reproduce
faithfully {\it all} the features of the relaxation dynamics of the
real material it can be expected that the salient properties are at least
correct from a qualitative point of view.

In the present simulation we integrated the equations of motions
using the velocity form of the Verlet algorithm with a time step
of 1.6~fs. At each temperature we first equilibrated the system by
coupling it to a stochastic heat bath. This equilibration time exceeded
the structural relaxation time of the silicon atoms (measured by means
of the incoherent intermediate scattering function for a wave--vector
1.7~\AA$^{-1}$, which corresponds to the length scale of the distance between
two tetrahedra, see below). Subsequently we started a production run
in the NVE ensemble using a density of 2.37~g/cm$^3$ which is close to
the experimental value~\cite{mazurin83}. The number of particles was
8064 ($N_{\rm Si}=1792$, $N_{\rm O}=4480$, $N_{\rm Na}=1792$) in a
cubic simulation box of size $L=48.653$~\AA. Such a large system size
is needed if one wants to avoid finite size effects in the relaxation
dynamics~\cite{horbach96,horbach01c}. In addition, previous simulations
have shown that sodium--silicate glasses have structural features that
extend over 6--8~\AA~\cite{horbach02,oviedo98,jund01} and therefore also
this calls for rather large system sizes.  The temperatures investigated
were 4000~K, 3400~K, 3000~K, 2750~K, 2500~K, 2300~K, and 2100~K. At the lowest
temperature the length of the production run was 1.5 million time steps,
which corresponds to 2.5~ns. In order to improve the statistics of the
results we did at each temperature two completely independent runs.

\section{Results}
\label{sec3}

The main quantities of interest in the present paper are $F({\bf q},t)$
and $F_s({\bf q},t)$, the coherent and incoherent intermediate scattering
functions for wave--vector ${\bf q}$. (Note that for isotropic systems,
like the one studied here, these space--time correlation functions
depend only on $q$, the modulus of ${\bf q}$, i.e.  there is no
directional dependence. Therefore we have made use of this fact also
in the analysis of our data and have averaged over all wave--vectors
with the same modulus.)  These two observables are not only of great
theoretical interest~\cite{hansen86} but can also be measured directly
in neutron--scattering experiments~\cite{lovesey}. They are given by

\begin{eqnarray}
F^{\alpha\beta}(q,t)& = &\frac{1}{N} \sum_{j=1}^{N_\alpha} \sum_{k=1}^{N_\beta}
\langle \exp\left[ i {\bf q} \cdot ({\bf r}_j(t)-{\bf r}_k(0))\right]\rangle \label{eq3}\\
F_s^\alpha(q,t) & = & \frac{1}{N} \sum_{j=1}^{N_\alpha}
\langle \exp\left[ i {\bf q} \cdot ({\bf r}_j(t)-{\bf r}_j(0))\right]\rangle \quad.
\label{eq4}
\end{eqnarray}

\noindent
Here ${\bf r}_j(t)$ is the position of particle $j$ at time $t$, $N_{\alpha}$
is the number of atoms of type $\alpha$, and $N$ is the total number of atoms. 

In Ref.~\cite{horbach01} we have shown that in NS2 the partial structure
factors $S_{\rm SiSi}(q)$, $S_{\rm SiO}(q)$, and $S_{\rm OO}(q)$ have
{\it two} pre--peaks: one at $q_1\equiv$ 0.94~\AA$^{-1}$ and a second
one at $q_2\equiv$ 1.7~\AA$^{-1}$. (The ``main''--peaks, corresponding
to the length scale of a nearest neighbor pair of Si--O and Na--O are
at $\approx 2.8$~\AA$^{-1}$ and $\approx 2.1$~\AA$^{-1}$, respectively.) 
The peak at $q_2$ is related to the distance between
neighboring tetrahedra and thus corresponds to the so--called ``first
sharp diffraction peak'' in pure silica. The peak at $q_1$ has recently
been shown to be related to the typical distance between the channels
mentioned in the introduction~\cite{horbach02}. Unfortunately, due
to cancellation effects in the partial structure factors (weighted
with the appropriate experimental neutron scattering cross sections)
it is hard to see this peak in a neutron scattering experiment done
at room temperature~\cite{meyer02,horbach01,misawa}. However, the recent
neutron--scattering studies of Meyer {\it et al.} have shown that this
system shows at high temperatures, i.e.~above $T\approx 1200$~K which
is well above the glass transition temperature~\cite{mazurin83}, a feature in the elastic
signal at a wave--vector around 0.9~\AA$^{-1}$, i.e.~very close to
$q_1$~\cite{meyer02}. Thus we conclude that this structural feature is
not only seen in the present model for NS2, but can be found in the real
material as well.  In the following we will demonstrate that these structural
features have also a strong influence on the relaxation dynamics of
the system.

\begin{figure}[h]
\hspace*{30mm}
\psfig{figure=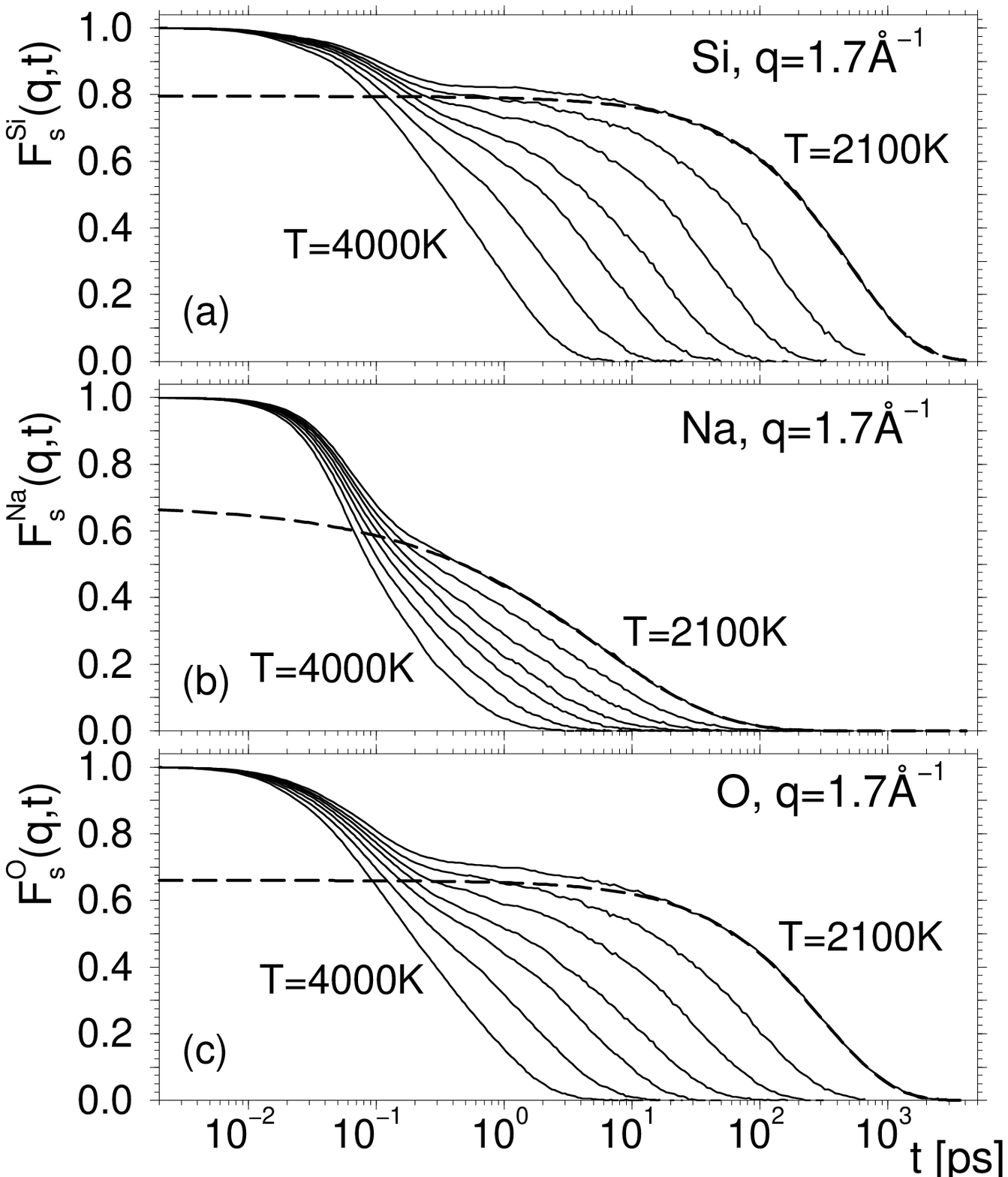,width=70mm,height=10.0cm}
\caption{Time dependence of the incoherent intermediate scattering
function for Si, O, and Na (top to bottom) for all temperatures
investigated and a wave--vector \protect$q=q_2=1.7$~\AA$^{-1}$. The dashed lines
are fits to the curve for \protect$T=2100$~K with a KWW-law.}
\label{fig1}
\end{figure}

In Fig.~\ref{fig1} we show the time dependence of the incoherent
intermediate scattering function for all temperatures investigated. We see
that at high temperatures the relaxation of the structure is relatively
fast in that the correlation function decays, after the microscopic time
scale which lasts around 0.2~ps, basically exponentially. From this graph
we also recognize that at these temperatures this decay is about a factor
of four faster for the sodium atoms than the one for the silicon atoms,
and about a factor of three faster than the one for the oxygen atoms. This
is in agreement with the values for the diffusion constants which at this
temperature show the same dependence on the species~\cite{horbach01}. At
low temperatures the correlators have a time dependence that differs
qualitatively from the one at high $T$ in that they show a plateau at
intermediate times. This plateau is related to the fact that on this
time scale the particles are trapped by their surrounding neighbors
(a behavior which is often called ``cage--effect'') and hence the time
correlation function changes only slowly. Only for relatively large
times the particles are able to leave this (temporary) cage and hence
the correlators finally decay to zero. It is customary to call the
dynamics in the time window in which the correlators are close to the
plateau the ``$\beta$--relaxation'' whereas the second relaxation step
is called the ``$\alpha$--relaxation''.  From the figure we see that at
low temperature, $\tau(q)$ (the time scale for the $\alpha$--relaxation)
of the sodium atoms is about a factor of 30 smaller than the one for
the oxygen atoms and about a factor of 100 smaller than the one for
silicon. Hence we conclude that the temperature dependence of $\tau$
depends on the species, also this in agreement with the one found for
the diffusion constant~\cite{horbach01}. Below we will discuss this
dependence in more detail.

Also included in the figure is a fit to the curves at the lowest
temperatures with a Kohlrausch--Williams--Watts--law (KWW), i.e.
$F_s^\alpha(q,t)=A\exp(-(t/\tau)^{\beta})$ (dashed lines). We see that
this functional form is able to describe the data very well in the
$\alpha$--relaxation regime, as it is usually the case for the relaxation
dynamics of glass--forming liquids. The value of the exponent $\beta$
is 0.8 for Si and O and 0.47 for Na. Thus we see that the relaxation
dynamics is not very stretched for the atoms making up the matrix whereas
it is very stretched for the network modifier. Below we will come back to
a possible interpretation for this low value of $\beta$. (Note that a
high value of $\beta$ for Si and O is in agreement with the observation
that also in pure silica the correlators for these length scales are not
very stretched~\cite{horbach01}. The presence of sodium does, however,
indeed lower the value of $\beta$ somewhat.)

\begin{figure}[h]
\hspace*{30mm}
\psfig{figure=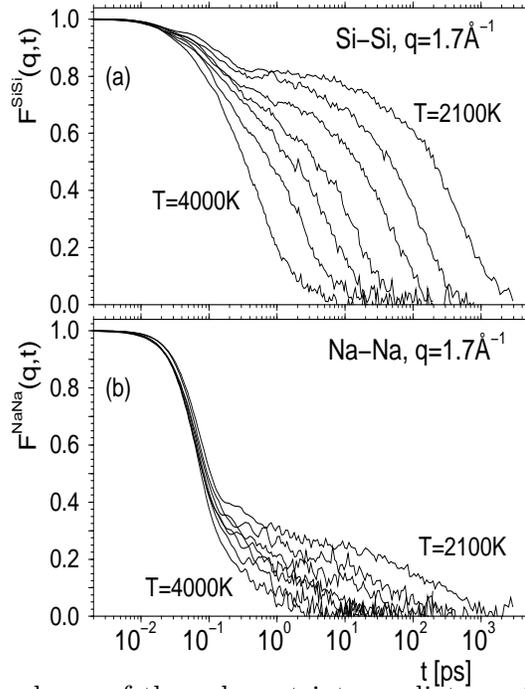,width=70mm,height=90mm}
\caption{Time dependence of the coherent intermediate scattering function for the
Si--Si, (a), and Na--Na, (b), correlation for all temperatures investigated.}
\label{fig2}
\end{figure}

Having discussed the species and temperature dependence
of $F_s^\alpha(q,t)$ we now focus on the coherent function
$F^{\alpha\beta}(q,t)$, defined in Eq.~(\ref{eq3}), since that one gives
the information how the overall structure of the system relaxes.  The time
and temperature dependence of $F^{\rm SiSi}(q,t)$ and $F^{\rm NaNa}(q,t)$
is shown in Fig.~\ref{fig2} (the one for the O--O correlation is similar
to the one for Si--Si). A comparison of the curves in this figure with
the incoherent functions, Fig.~\ref{fig1}, shows that for the case of
silicon the coherent and incoherent functions are very similar in that
the height of the plateau at intermediate times as well as the typical
relaxation times for the $\alpha$--relaxation are quite comparable. This is
the usual behavior found for glass forming systems such as pure silica,
water, or simple liquids~\cite{horbach01b,sciortino,kob95}. In contrast
to this the time dependence of $F_s^{\rm Na}(q,t)$ for sodium differs
strongly for the coherent function $F^{\rm NaNa}(q,t)$ in that the
relaxation time of the former is about a factor of ten smaller than
the one of the latter. This shows that the relaxation dynamics of
the sodium atoms has unusual features. In fact it has been shown in
Refs.~\cite{horbach02,jund01} that the motion of a tagged Na atom is
relatively fast in that the particles hop between certain preferential
sites (generated by the Si--O matrix). However, the {\it relative} spatial
arrangement of the Na particles (i.e.~their global structure) does not
change under this type of motion and hence the coherent function decays
much slower than the incoherent one.

\begin{figure}[h]
\hspace*{30mm}
\psfig{figure=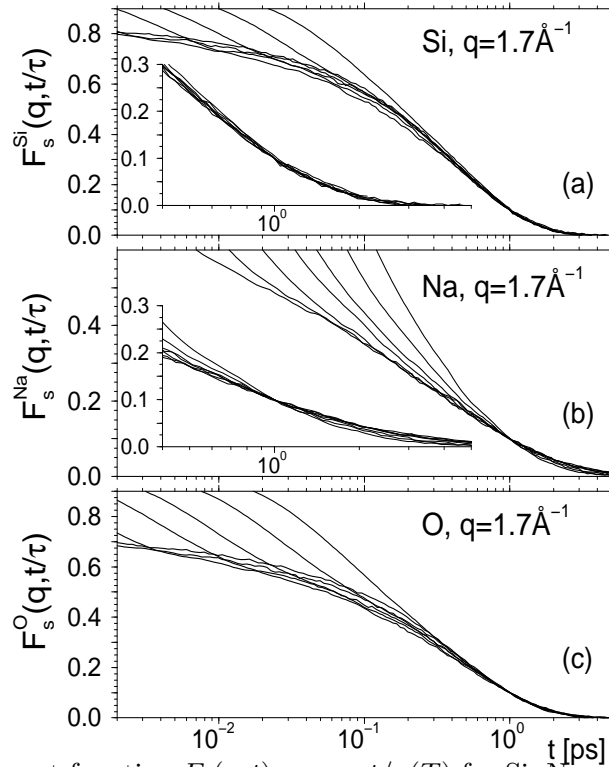,width=8.2cm,height=10.0cm}
\caption{The incoherent function \protect$F_s(q,t)$ versus
\protect$t/\tau(T)$ for Si, Na, and O (top to bottom). The
\protect$\alpha$--relaxation time \protect$\tau(T)$ has been defined
via \protect$F_s(q,\tau)=0.1$. The inset in (a) and (b) show the same
functions at large rescaled times.}
\label{fig3}
\end{figure}

One of the important predictions of MCT is that close to the critical
temperature of the theory the shape of the time correlation functions
does not depend on temperature. This means that a correlator $\phi(t,T)$
can be written as

\begin{equation}
\phi(t,T)=\hat{\phi}(t/\tau(T)) \quad,
\label{eq5}
\end{equation}

\noindent
where $\tau(T)$ is the $\alpha$--relaxation time at temperature
$T$.  To what extent this prediction, which is often called the
time--temperature superposition principle (TTSP), holds can easily be
tested by plotting the correlators versus $t/\tau$. For this we have
defined the $\alpha$--relaxation time $\tau$ as the time it takes the
correlator to decay to 0.1. Although this definition (and the value 0.1)
is somewhat arbitrary, it is easy to see that {\it if the TTSP holds} the
details of the definition do not matter. In Fig.~\ref{fig3} we show the
correlators as a function of $t/\tau$ for all temperatures investigated.
We see that for silicon and oxygen the TTSP holds basically for the
whole $T$--range. In the case of sodium we have already shown that its
dynamics is very different from that of Si and O in that, e.g., for
$F_s^{\rm Na}(q,t)$ one obtains a very different stretching exponent
$\beta$ from that for $F_s^{\rm Si}$ and $F_s^{\rm O}$ at a given $q$.
However also for $F_s^{\rm Na}(q,t)$ the TTSP work for low temperatures,
$T \le 2500$~K. That for this species the TTSP does not work for
higher temperatures is reasonable since for $T\ge 2750$~K the typical
$\alpha$--relaxation times are still of the order of a few picoseconds
only (see Fig.~\ref{fig1}b) and thus the $\alpha$--relaxation regime
is not well separated from the microscopic dynamics. Note that if one
would be able to equilibrate the system at even lower temperatures it
can be expected that the TTSP starts to break down, since at low $T$
the dynamics is likely to be dominated by a simple diffusive motion
in which the single step is the breaking of a Si--O bond, a behavior
which has been found in {\it pure} silica~\cite{horbach99}. Hence one
can expect that this affects also the sodium dynamics and thus to the
breaking down of the TTSP.

\begin{figure}[h]
\hspace*{15mm}
\psfig{figure=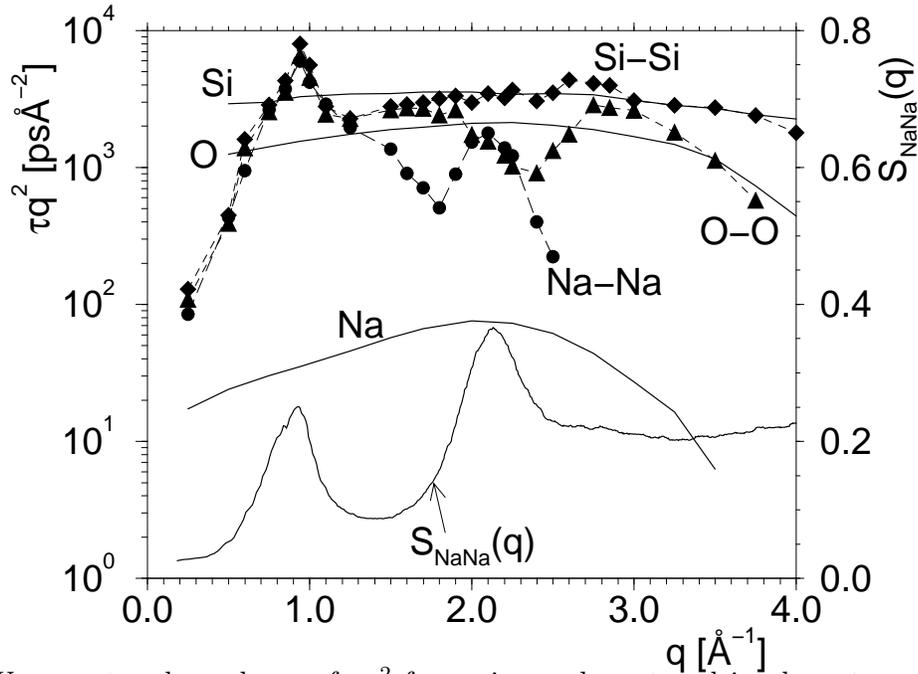,width=120mm,height=90mm}
\caption{Wave--vector dependence of \protect$\tau q^2$ for various
coherent and incoherent correlators (see labels of the curves) at $T=2100$~K.
Right scale: \protect$q$--dependence of the partial structure
factor for the sodium atoms.}
\label{fig4}
\end{figure}

The correlators discussed so far were for $q=1.7$~\AA$^{-1}$, the length
scale corresponding to the typical distance between two neighboring
tetrahedra. We now investigate how the $\alpha$--relaxation time depends
on the wave--vector as well as on the particle species. In Fig.~\ref{fig4}
we show the $q$--dependence of $\tau$ as determined from the incoherent
correlators as well as some of the coherent ones.  The temperature
is $T=2100$~K, thus the lowest temperature at which we were able to
fully equilibrate the system. Since for the case of a diffusive motion
the relaxation time $\tau(q)$ of the incoherent function is equal to
$(D q^2)^{-1}$ ($D$: diffusion constant) we plot directly $\tau(q)q^2$
which can be understood as an inverse $q$--dependent diffusion constant.
We first start with the particles forming the matrix, i.e.~Si and O. We
see that for these elements $\tau(q)q^2$ from the incoherent functions
increases weakly with increasing $q$, goes through a maximum and then
decreases again at large $q$. The location of this maximum is around
2.8~\AA$^{-1}$, i.e.~at the wave--vector at which the static structure
factor has the peak that corresponds to the distance between neighboring
Si--O.  The same type of $q$--dependence has also been found in the
case of pure SiO$_2$~\cite{horbach01b} and it reflects the fact that
on this length scale the system is particularly rigid. The relaxation
times of the coherent functions have a more complex $q$--dependence in
that they oscillate around the $\tau$ for the incoherent function. This
oscillation is in phase with the corresponding static structure factor,
an effect that is known as de Gennes--narrowing~\cite{hansen86} and which
is found in simple liquids~\cite{kob95} but also in a system like pure
silica~\cite{horbach01b}, or water~\cite{sciortino}. From the figure
we recognize that the largest relaxation time is at a wave--vector
$q\approx 0.95$~\AA$^{-1}$, which corresponds to the length scale of
the network of channels discussed in Refs.~\cite{horbach02,jund01}.
The different chemical ordering of silicon and sodium leads to the
presence of an additional intermediate length scale which is reflected
in the dynamics of sodium in that the trajectories of the sodium atoms
are restricted to a network of channels in a Si--O matrix. This network
of channels is reflected in the collective correlations by the slowest
relaxation process of our system, the rearrangement of the channel
structure. Finally we remark that for wave--vectors below 0.95~\AA$^{-1}$
$\tau(q)q^2$ exhibits a very steep increase: We see that in this range
of $q$ values the curves are very well compatible with a straight line
which corresponds to a growth of the relaxation times like $\tau(q)
\propto q^{-2}\exp(Aq)$, where $A$ is a positive constant. Although we
are not aware of any theoretical reason for such a dependence it seems
to describe our data remarkably well over two decades in $\tau$.

For the sodium atoms the $q$--dependence of $\tau$ is more complicated
than the one for Si and O.  For the coherent correlators this function is
qualitatively similar to the one found for Si--Si and O--O. In particular
we find again a pronounced peak at $q\approx 0.95$~\AA$^{-1}$, i.e.~the
length scale of the channels. Note that close to this peak also the
absolute value of $\tau$ is close to the ones for Si and O which shows
that on this length scale the spatial arrangement of the Na atoms can
only relax if the Si--O matrix relaxes. The relaxation time for the
incoherent Na function behaves very differently. First of all we see that
it is significantly smaller than the one of the coherent function, in
agreement with our conclusions from Figs.~\ref{fig1} and \ref{fig2}. Also
this function increases for small and intermediate wave--vectors, shows
a maximum at the location of the corresponding peak in the partial
structure factor (included in the figure as well) and then decreases
for even larger $q$.  It is not possible to determine the relaxation
times of $F^{\rm NaNa}(q,t)$ for $q>2.5$~\AA$^{-1}$ with our definition
$F^{\alpha \beta}(q, t=\tau)=0.1$ because for large wave-vectors $F^{\rm
NaNa}(q,t)$ has already decayed to values around or lower $0.1$ before the
$\alpha$--relaxation starts to develop (see Fig.~\ref{fig8} below). In
contrast to that we can determine the relaxation times for $F_s^{\rm
Na}(q,t)$ also for relatively large wave--vectors. Since for sufficiently
large $q$ one has $F_s^\alpha(q,t) \approx F^{\alpha\alpha}(q,t)$, we know
that in this limit the relaxation times for $F^{\alpha\alpha}(q,t)$ have
to approach the ones for $F_s^\alpha(q,t)$. This implies that the curve
$\tau(q)$ for the Na--Na correlation must start to decrease quickly in
order to approach the one for the Na correlation. Hence we expect that
for large wave--vectors the relaxation times for the Na--Na correlation
become significantly smaller than the one of the matrix. An inspection
of the correlation functions shows however, that this is not yet the
case for wave--vectors smaller than 3.5~\AA$^{-1}$.

In an earlier study of pure silica, the paradigm of a ``strong''
glass--former, we found the surprising result that the temperature
dependence of the diffusion constant as well as the viscosity show at
high temperatures a significant deviation from the expected Arrhenius
law~\cite{horbach99}. In that paper it was argued that this deviation can
be rationalized by mode--coupling theory which predicts the existence of a
``critical temperature'' $T_c$ close to which the transport coefficients
show a non--Arrhenius behavior~\cite{mct}. Since significant deviations
from an Arrhenius law have been observed also for the sodium--silicate
system investigated here~\cite{horbach01} it is reasonable to see to
what extent MCT is able to rationalize the relaxation dynamics. The
theory predicts that close to $T_c$ the temperature dependence of the
diffusion constants or $\alpha$--relaxation times is given by a power--law:

\begin{equation}
D(T) \propto (T-T_c)^\gamma \quad {\rm and} \quad \tau(T) \propto (T-T_c)^{-\gamma}.
\label{eq6}
\end{equation}

\begin{figure}[h]
\hspace*{15mm}
\psfig{figure=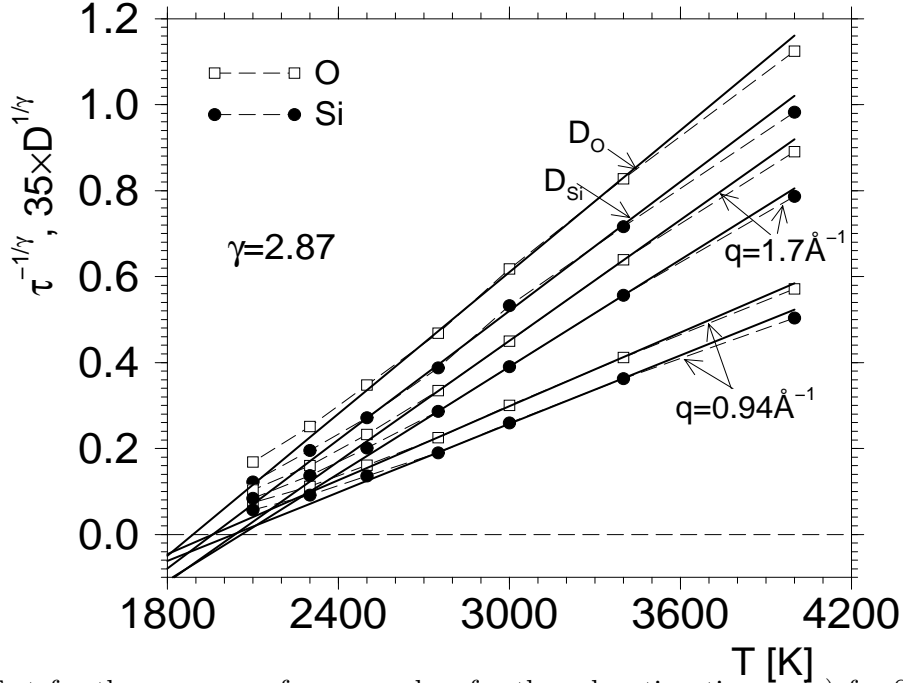,width=120mm,height=90mm}
\caption{Test for the presence of a power--law for the relaxation times
\protect$\tau(q)$ for Si and O and two wave--vectors. Also included is the
data for the diffusion constants \protect$D_{\rm Si}$ and \protect$D_{\rm
O}$, which for the sake of a clearer presentation have been multiplied by
35. The bold straight lines are linear fits corresponding to the theoretical
expectation.}
\label{fig5}
\end{figure}

\noindent
Here $\gamma$ is a system--universal constant, i.e.~it does not depend on what
species or wave--vector one considers. In addition MCT predicts that the
value of the exponent $\gamma$ has a one--to--one correspondence with the
exponent $b$ of the so-called von Schweidler-law that is discussed below. This
connection is given by

\begin{equation}
\gamma=\frac{1}{2a}+\frac{1}{2b} \quad {\rm with}\quad 
\frac{[\Gamma(1-a)]^2}{\Gamma(1-2a)}= \frac{[\Gamma(1+2b)]^2}{\Gamma(1+2b)} \quad ,
\label{eq7}
\end{equation}

\noindent
i.e.~the second equation can be used to determine the value of $a$ from
$b$ and then the first equation can be used to calculate $\gamma$. (Here
$\Gamma(x)$ is the usual $\Gamma-$function.) In Ref.~\cite{horbach02} we
have shown that the value of $b$ is around 0.47 (see also Fig.~\ref{fig6}
below). Hence Eqs.~(\ref{eq7}) give a value of $\gamma=2.87$, which can
be considered as the theoretical estimate of MCT for the exponent. If
the prediction of the theory on the power--law and the value of the
exponent is correct, a plot of $\tau^{-1/\gamma}$ (or $D^{1/\gamma}$)
vs. $T$ should give a straight line. This type of plot is shown in
Fig.~\ref{fig5} for the relaxation times $\tau(q)$ for the wave--vectors
$q=0.94$~\AA$^{-1}$ and $q=1.7$~\AA$^{-1}$ as well as the diffusion
constants (which were determined from the long time limit of the mean
squared displacement of a tagged particle~\cite{horbach01}). Note
that we have not included the data for Na, since, as shown in
Ref.~\cite{horbach01}, the diffusion constant for Na follows an Arrhenius
law in the whole temperature range, and hence the power--laws given by
Eq.~(\ref{eq6}) certainly do not hold for this species. We see that all
the curves do indeed show a straight line in a temperature interval that
is quite substantial. Linear fits in this region are included in the graph
as well (solid lines).  Furthermore we see that the extrapolation of these
straight lines to lower temperatures intersects the $T$--axis at a point,
the critical temperature $T_c$, which depends only weakly on the time
considered, and which is around 2000~K. Hence we conclude from this figure
that the relaxation times and diffusion constants for Si and O do indeed
show the predicted power--law dependence with a common exponent $\gamma$
and a common critical temperature $T_c$. Finally we note that in the
temperature interval in which $\tau$ and $D$ show this power--law these
quantities change by about two orders of magnitude and hence the existence
of the power--laws is not just in a trivial matter. Note that although
MCT predicts that at $T_c$ the relaxation times should diverge, in reality
this is not found.  The reason for this is that once the relaxation times
have increased beyond a certain value, for atomic systems usually on the
order of 10~ns, the system starts to relax via processes that currently
can be taken into account by the theory only in a schematic way.  Despite
the presence of these processes, usually called ``hopping processes''
the theory is still able to make prediction on the relaxation dynamics
on the time scale of the $\beta$--relaxation. For more details we refer
to Refs.~\cite{mct,hopping}. The presence of these hopping processes
for the present system can be inferred from Fig.~\ref{fig5} in that the
data points for $T\leq 2600$~K are above the theoretical straight line,
i.e. the relaxation is faster than predicted from the power--law.

Having discussed the relaxation dynamics of the system in the
$\alpha$--regime, we now turn our attention to the $\beta$--regime,
i.e.~the time window in which the correlators are close to the plateau
(see Fig.~\ref{fig1}). One of the main predictions of MCT is that in the
$\beta$--regime and for temperatures close to $T_c$ the time dependence of
the correlation function is system universal in that any time correlation
function $\phi(t)$ can be written as

\begin{equation}
\phi(t)= \phi_c+h_{\phi} G(t) \quad .
\label{eq8}
\end{equation}

\noindent
Here $\phi_c$ is the height of the plateau, often also called
non--ergodicity parameter, and the whole time dependence is in the
system universal function $G(t)$. Due to its structure, Eq.~(\ref{eq8})
is often called ``factorization property'', since $\phi(t)-\phi_c$
factors into a time dependent function and a $\phi$--dependent function.
(Note that this factorization property holds also for the case that
hopping processes are present~\cite{mct,hopping}.) The time dependence of
$G(t)$ is given by the solution of a non--linear equation which can be
solved numerically~\cite{mct}. However, it can be shown that very close to
$T_c$ this solution is given by the following form~\cite{bgs84}:

\begin{equation}
G(t)= -Bt^b+B't^{2b} \quad .
\label{eq9}
\end{equation}

\noindent
This approximation is good for times at which the correlators have
started to fall below the plateau, but are still close to it. The first
power--law of Eq.~(\ref{eq9}) is often called ``von Schweidler--law''
and the exponent $b$ the ``von Schweidler--exponent''. Note that,
since $G(t)$ is predicted to be independent of the correlator,
also the value of $b$ should be the same for all $\phi(t)$. However,
since Eq.~(\ref{eq9}) also includes the first correction term to this
asymptotic result, the coefficient $B'$ will depend on the correlator. In
Fig.~\ref{fig6} we show the time dependence of the incoherent intermediate
scattering function for the oxygen atoms for various wave--vectors (at
$T=2100$~K). Also included are fits with the functional form given by
Eqs.~(\ref{eq8}) and (\ref{eq9}) using the height of the plateau as a
fit parameter. The dotted lines correspond to a fit in which $G(t)$ is
only given by the von Schweidler--law, i.e. the first term on the RHS of
Eq.~(\ref{eq9}), whereas the dashed line is the case that also the second
term in Eq.~(\ref{eq9}) is taken into account. In these fits $B$ and
$B'$ were fit--parameters that were allowed to depend on $q$, whereas the
exponent $b$ was a global fit--parameter. From the figure we conclude that
the von Schweidler--law is indeed able to describe well the dynamics close
to the plateau. Furthermore we see that the inclusion of the correction
term increases the time window for which this law holds by about a factor
of ten in qualitative agreement with the results of such an analysis for
other glass--forming liquids~\cite{mct,horbach01b,sciortino,kob95}. We
also mention that a similar good fit is obtained for the case of
$F_s(q,t)$ for Si.  In Ref.~\cite{horbach02} we showed the same type
of fits for the coherent functions of Si and O and found that also for
these correlators the $\beta$--regime is described very well by the
functional form given in Eq.~(\ref{eq9}). Hence we conclude that in
the $\beta$--regime the relaxation dynamics is indeed independent of
the correlator.

We emphasize that this universality holds only for the $\beta$--relaxation
regime and not for the $\alpha$--relaxation. For the latter one finds,
that the stretching exponent $\beta$, and therefore the shape of the
correlator, depends on the species or the wave--vector. In particular we
have discussed in Ref.~\cite{horbach02} the wave--vector dependence of
$\beta$ for the Na atoms and have shown that for small and intermediate
$q$, $q\leq 1.5$~\AA$^{-1}$, $\beta$ changes significantly thus showing
that the $\alpha$--relaxation is indeed not universal from this point
of view.

In this context it is, however, very interesting that
$\beta(q)$ for the Na atoms becomes independent of $q$ for $q\geq
1.6$~\AA$^{-1}$~\cite{horbach02}, {\it although} in this $q$--range the
structure factor $S_{\rm NaNa}(q)$ still shows pronounced features,
i.e. is not a constant. Such a behavior was predicted some time ago
by Fuchs who used MCT to show that $\lim_{q\to \infty} \beta(q)=b$,
i.e.~for large wave--vectors $\beta$ should converge to the von
Schweidler exponent $b$~\cite{fuchs94}. We have found that in our case
this is indeed the case, i.e.~that for large $q$ the stretching exponent
is indeed compatible with $b=0.47$~\cite{horbach02}. Hence this nice
agreement between our results and the prediction of MCT shows that the
theory is indeed able to describe also this features of the relaxation
dynamics of the present system. We also mention that for the case of a
hard sphere system the MCT prediction for the independence of $\beta$
of $q$ holds only for wave--vectors that are much larger than the
location of the first peak in the structure factor. The fact that for
the present system this asymptotic value is reached already for quite
small wave--vectors is thus rather surprising. We note, however, that a
closer inspection of $F_s^{{\rm Na}}(q,t)$ for $q\geq 1.6$~\AA$^{-1}$
shows that the $\alpha$--relaxation regime of these correlators fall
into the late $\beta$--relaxation regime of all the other (slow)
correlators in which the von Schweidler law holds. Thus, it seems that
the relaxation processes in the $\beta$--regime of the slow correlators
that correspond to the universal von Schweidler decay are impressed onto
the dynamical behavior of the single particle motion of the sodium atoms
which leads to stretched exponential decay of $F_s^{{\rm Na}}(q,t)$ with
$\beta=b$. For smaller wave-vectors this is not the case since there the
$\alpha$--relaxation of $F_s^{{\rm Na}}(q,t)$ overlaps essentially only
with the {\it plateau} of the slow correlators.

The result in Fig.~\ref{fig6} shows that the time dependence of
the correlation functions for the time regime in which they start to
fall below the plateau is compatible with the functional form given in
Eq.~(\ref{eq9}). 
The factorization property stated in Eq.~(\ref{eq8})
is, however, more general, since expression~(\ref{eq9}) is just the
leading asymptotic prediction for $G(t)$. A different way to check to
what extent the factorization property holds, without making use of the
explicit form of $G(t)$, is to calculate the following quantity:

\begin{equation}
R_{\phi}(t)=\frac{\phi(t)-\phi(t')}{\phi(t'')-\phi(t')} \quad .
\label{eq10}
\end{equation}

\noindent
Here $t'$ and $t''$ are two arbitrary times in the $\beta$--regime. It
follows immediately that {\it if} Eq.~(\ref{eq8}) holds, $R_{\phi}$ is
independent of $\phi$, since it is just the system universal function
$G(t)$. In Fig.~\ref{fig7} we show the time dependence of $R_{\phi}$
at $T=2100$~K. The correlators $\phi$ used are $F_s(q,t)$ for Si and
O at $q=0.94$~\AA$^{-1}$, $q=1.7$~\AA$^{-1}$, $q=2.0$~\AA$^{-1}$,
and $q=3.0$~\AA$^{-1}$, as well as the coherent functions $F(q,t)$
for Si--Si, Na--Na, and O--O at the same values of $q$. The times $t''$
and $t'$ from Eq.~(\ref{eq10}) are 2.9~ps and 10.3~ps, respectively. From
the figure we recognize that in the $\beta$--regime the $R_{\phi}(t)$ for
all these correlators collapse nicely onto a master function, which is
the function $G(t)$. Hence we conclude that the factorization property
predicted by MCT holds for the present system. Finally we mention that
we find for higher temperatures the same type of collapse, but that
the time window in which the master curve is observed shrinks rapidly,
in qualitative agreement with the prediction of MCT.

The last quantity we will discuss is $\phi_c$ from Eq.~(\ref{eq8}). Since
$\phi_c$ is just the height of the plateau of the correlator at
intermediate times, this parameter is often also called ``non--ergodicity
parameter'' (NEP) since it reflects how much memory the system has of
its state at $t=0$. In the following we will focus on the wave--vector
dependence of $f^{\alpha\beta}(q)$ and $f_s^\alpha(q)$, the NEP
for the coherent and incoherent scattering functions.  (Note that
$f^{\alpha\beta}(q)$ and $f_s^\alpha(q)$ are often also called the
Debye--Waller factor and Lamb--M\"ossbauer--factor, respectively.)
We have determined $f^{\alpha\beta}(q)$ and $f_s^\alpha(q)$ by
using Eq.~(\ref{eq8}) to fit the correlators. The $q$--dependence
of $f^{\alpha\beta}(q)$ and $f_s^\alpha(q)$ for the case of silicon
and oxygen are shown in Fig.~\ref{fig8}. We see that in both cases
$f_s^\alpha(q)$ (open symbols) decays quickly with increasing $q$ and
that this dependence can be described very well by a Gaussian (bold
solid lines). Such a behavior has already been found for the case of
pure silica~\cite{horbach01} and is in qualitative agreement with the
prediction of MCT~\cite{mct}. The width $q_s$ of these Gaussians are
4.1~\AA$^{-1}$ and 3.0~\AA$^{-1}$ for Si and O, respectively. This
means that in the time scale of the $\beta$--relaxation the particles
are trapped in a cage with radius $r_s=1/q_s=0.24$~\AA~(Si) and
0.33~\AA~(O). For the case of pure SiO$_2$ the corresponding values are
0.23~\AA~and 0.29~\AA~\cite{horbach01}. Hence we see that the presence of
sodium slightly increases the size of the cage and that this increase is
more pronounced for the case of oxygen than for silicon. This result is
reasonable since some of the oxygen atoms are in the immediate vicinity
of the sodium atoms but are bound to them less strongly than they are to
the silicon atoms (and of course there is no Si--Na bond). In addition
also the presence of dangling bonds (i.e. non--bridging oxygens) will
lead to an increase of the size of the cage for oxygen.

The NEP for the coherent functions of Si and O oscillate around the
ones for the incoherent functions. This oscillation is in phase with
the corresponding structure factor, a behavior which is in qualitative
agreement with the theoretical expectation~\cite{mct}. In particular
we see that the amplitude of this oscillation is smaller for the case
of silicon than for oxygen, in agreement with the findings of pure
SiO$_2$~\cite{horbach01b} or binary mixtures of particles~\cite{kob95}. This
finding can be rationalized by the fact that in a binary system with
strong asymmetry in the concentration the coherent correlation functions
for the minority species are in general very similar to the incoherent
functions.

Also included in the figure is $f^\alpha(q)$ as determined from
the Na correlator. (Note that $F_s^{\rm Na}(q,t)$ does not show a
well defined plateau for $q\ge 1.5$~\AA$^{-1}$ (see e.g.~Fig.~2 in
Ref.~\cite{horbach02}). Hence a fit with the functional form given
by Eq.~(\ref{eq8}) is rather difficult if $q$ is large and thus the
height of the plateau cannot be determined with high accuracy. However,
for $q\leq 1.2$~\AA$^{-1}$ one does indeed find a well developed
plateau and thus it is possible to determine $f_s^\alpha(q)$ with good
accuracy.) From Fig.~\ref{fig8} we see that $f_s^\alpha(q)$ for Na decays
significantly faster than the ones for Si and O. This can be interpreted
by saying that the cage for the sodium atoms is wider than the one
for Si and O. However, care must be taken in drawing this conclusion
since, as just mentioned, at the temperatures investigated neither
the intermediate scattering functions for intermediate and large $q$
nor the mean--squared displacement of the Na atoms show a well defined
plateau at intermediate times. Hence one cannot really say that on this
time scale the particles are caged and hence also the interpretation of
$f_s^\alpha(q)$ as the Fourier--transform of the shape of the cage is
not quite appropriate. What is remarkable with $f_s^\alpha(q)$ for Na is
the fact that it is {\it not} possible to fit it well with a Gaussian. If
one fits the data for $q\geq 2.2$~\AA$^{-1}$ with such a functional form
it is possible to obtain a very good fit (bold solid line). (The width
of this Gaussian is 2.52~\AA$^{-1}$, which corresponds to a ``cage''
of size 0.39~\AA.) However, this fit gives a very poor representation of
the data for smaller wave--vectors. This result, which is in contrast to
the findings for Si and O, shows that the dynamics of Na is indeed rather
unusual. Roughly speaking one thus can say that for small length scales,
i.e.~large $q$, the cage is relatively soft and therefore the $f_s^{\rm
Na}(q)$ decays quickly. However, on the length scales of the typical
distances between the channels (small $q$) the ``cage'' is relatively
rigid, since on this length scale, as we have mentioned before, the single
particle dynamics of Na is not strongly coupled to relaxation processes
in the matrix.  But of course this is just a hand--waving explanation
of the finding and it would be nice to find support for it by means of
a more thorough theoretical calculation.

Also included in the figure is the NEP for the Na--Na correlation.
Qualitatively this $f^{\alpha\beta}(q)$ looks similar to the one for the Si--Si
or O--O correlation and in particular it shows a pronounced peak at
around 2.0~\AA$^{-1}$, which corresponds to the nearest neighbor distance
between two Na atoms (=~3.3~\AA)~\cite{horbach01}. However, we notice an important
difference in that this NEP does not oscillate around the NEP for the
incoherent function but instead stays systematically below it. Also
for this behavior we are not aware of any theoretical prediction or
experimental result.

\section{Summary}
In this paper we have studied by means of molecular dynamics computer
simulations the relaxation dynamics of a melt of Na$_2$O--2SiO$_2$, one
of the prototypes of an ion--conducting glass former. In particular we
investigated the temperature and wave--vector dependence of the coherent
and incoherent scattering functions. Due to the very different time scale
of the dynamics of Na from the one of the species forming the matrix, the
$q$--dependence of these correlators for the sodium atoms shows features
that are neither found in simple liquids nor in network--forming liquids
like pure silica, i.e. systems in which the dynamics of the individual
species takes place on a comparable time scale. E.g.~we find that the
time and temperature dependence of the incoherent function for Na is
very different from the one for the coherent function. This is related
to the fact that in this system the sodium atoms have the tendency to
populate a relatively small subregion of space, so--called channels,
and that the dynamics of the atoms in these channels is relatively quick
and occurs by (activated) single particle hops. In contrast to this the
overall structure of the channel, and hence the coherent function for
the Na atoms, relaxes only on the time scale of the $\alpha$--relaxation
time of the Si--O matrix. The temperature dependence of the relaxation
times for the matrix and $F^{\rm NaNa}(q,t)$ show a strong deviation
from an Arrhenius law in agreement with experimental findings for this
system~\cite{hess96,knoche94}. In the past such deviations have been found
for simple liquids~\cite{kob95,hansen,hiwatari,kob} and their existence
has been rationalized by means of MCT. However, finding them in systems
like silica~\cite{horbach99,saika01} or in the present sodium silicate
system is rather surprising. In this paper we have shown that in the
temperature regime where these deviations are seen many of the features
of the relaxation dynamics can again be rationalized by means of MCT,
thus showing that with respect to this there is no difference with the
results found for the so--called fragile glass--formers. Hence we conclude
that the main difference between strong and fragile glass--formers is
that the presence of the hopping processes leads to a shrinking of the
dynamical range in which the $T$--dependence of the relaxation times
follows the power--law predicted by MCT (as compared to the range found
in fragile systems). Nevertheless, despite this reduced range, the time
correlation functions still show a behavior that can be rationalized
remarkably well by the theory.

Of course one has to wonder to what extent the results presented in
this paper can be found also in real Na$_2$O--2SiO$_2$, or similar
systems. Although it must be expected that the potential used is not
sufficiently accurate to reproduce all the properties of the real material
on a quantitative level, the surprisingly good agreement of the results
of the present model with the neutron--scattering results of Meyer {\it et
al.}~\cite{meyer02} shows that the potential is quite realistic. Therefore
it can be hoped that the results presented here will be found also in
appropriate coherent and incoherent neutron--scattering experiments
and we hope that the present work helps to motivate such experiments.

Acknowledgments: We thank A. Meyer for useful discussions. Part of this
work was made possible by the DFG through SFB 262 and Schwerpunktsprogramm
1055. We thank the HLRZ Stuttgart for a generous grant of computer time
on the CRAY T3E.

\vspace*{-6mm}

\vspace*{-6mm}

\begin{figure}[h]
\hspace*{10mm}
\psfig{figure=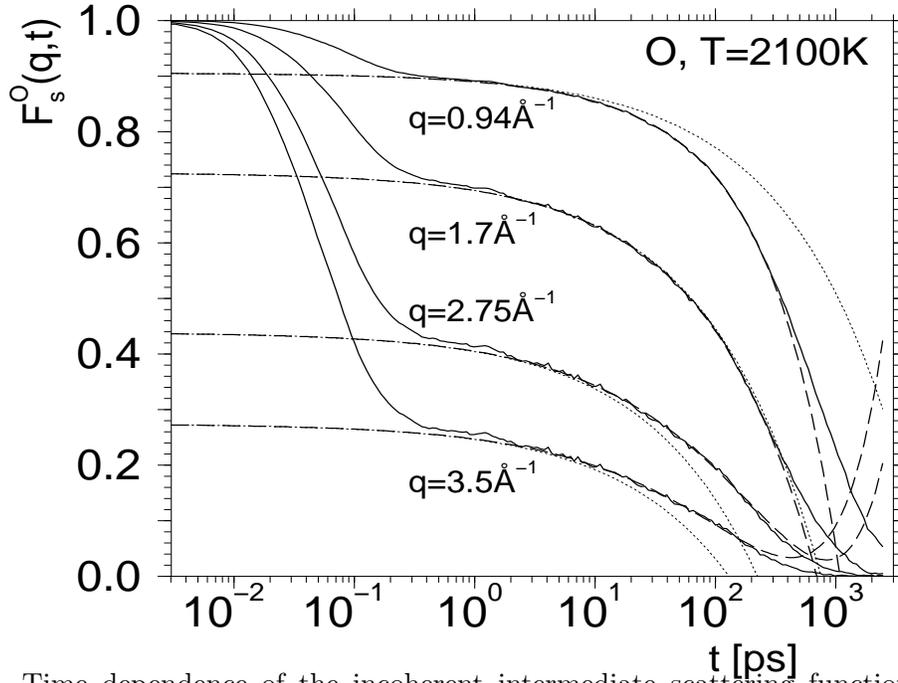,width=120mm,height=90mm}
\caption{Time dependence of the incoherent intermediate scattering
function for oxygen at \protect$T=2100$~K. The solid curves
are \protect$F_s^{\rm O}(q,t)$ for different wave--vectors (see
labels). The dotted curves are fits with a von Schweidler law, first
term in Eq.~(\protect\ref{eq9}), with an exponent \protect$b=0.47$. The
dashed curves are fits with the von Schweidler law including the leading
order corrections (see Eq.~(\protect\ref{eq9})).  }
\label{fig6}
\end{figure}

\begin{figure}[h]
\hspace*{15mm}
\psfig{figure=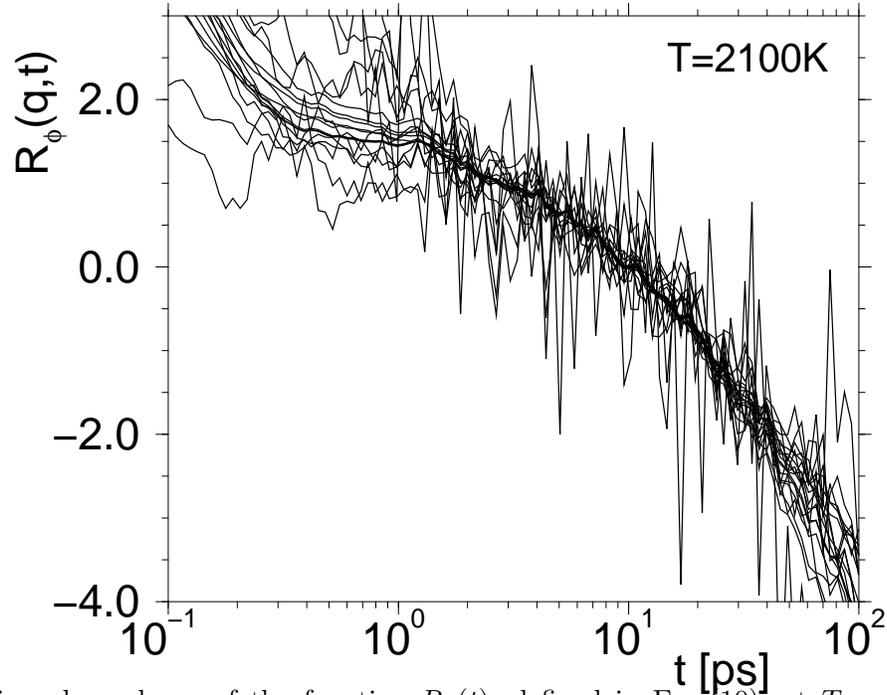,width=120mm,height=90mm}
\caption{Time dependence of the function \protect$R_\phi(t)$, defined
in Eq.~(\protect\ref{eq10}), at $T=2100$~K. The correlators used to make
this plot are \protect$F_s^{\rm Si}(q,t)$ and \protect$F_s^{\rm O}(q,t)$
for \protect$q=0.94, 1.7, 2.0, 3.0$~\AA$^{-1}$ and \protect$F^{\alpha
\alpha}(q,t)$ with \protect$\alpha \in \{ {\rm Si, O, Na} \}$ and at
the same wave--vectors.
}
\label{fig7}
\end{figure}

\begin{figure}[h]
\hspace*{15mm}
\psfig{figure=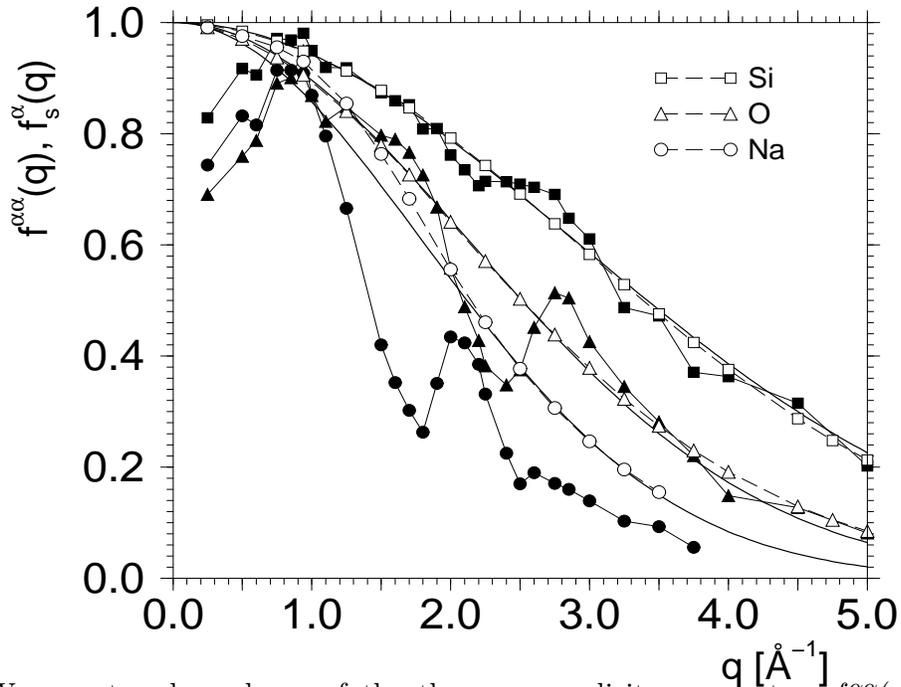,width=120mm,height=90mm}
\caption{Wave--vector dependence of the the non--ergodicity parameters
\protect$f^{\alpha\alpha}(q)$ and \protect$f_s^\alpha(q)$ (curves with
open and closed symbols, respectively) for \protect$\alpha \in \{{\rm Si,
O, Na}\}$. The bold lines are Gaussian fits to the NEP for the incoherent
functions.}
\label{fig8}
\end{figure}

\end{document}